\documentclass{iaus260}
\usepackage{graphicx}
\pubyear{2009}
\volume{260}  
\pagerange{1--4}
\jname{The R\^ole of Astronomy in Society and Culture}
\editors{D. Valls-Gabaud \& A. Boksenberg, eds.}

\title[Astronomy rare books exhibit at the National Library of Argentina]                   
{The National Library of Argentina: exhibiting astronomy--related rare books} 

\author[Alejandro Gangui]           
{Alejandro Gangui$^{1,2}$}        

\affiliation{$^1$Instituto de Astronom{\'\i}a y F{\'\i}sica del Espacio / CONICET, \\ 
                 Ciudad Universitaria, 1428 Buenos Aires, Argentina. \\ 
                 email: {\tt gangui@df.uba.ar} \\[\affilskip]
             $^2$Centro de Formaci\'on e Investigaci\'on en la Ense\~nanza de las Ciencias, \\ FCEyN, Universidad de Buenos Aires.}

\begin{document}
\maketitle

\begin{abstract}
Astronomical and cosmological knowledge up to the dawn of modern science was profoundly embedded in myth, religion and superstition. Many
of these inventions of the human mind remain today stored in different supports: medieval engravings, illuminated manuscripts, and of
course also in old and rare books.

\end{abstract}

\firstsection 
\section{Introduction}

Old and rare books: vestiges of the past which are well preserved in special reserves of main libraries, and constitute a source of pride
of these institutions. 

Among the many incunabula owned by Argentina's National Library, a couple of volumes stand out. The Liber Chronicarum cum figuris et
ymaginibus, compiled by the humanist Hartmann Schedel and printed by Anton Koberger in 1493 (the famous N\"urnberg chronicle) is one of the
``must see'' of the Library. It includes not only a description of Pliny's marvelous hominids -headless Libyans endowed with eyes and
mouth in their chests, Cyclopes from India and other {\it mirabilia}- but also many images with clear cosmological flavor. 

\begin{figure}[h!tb]
\begin{center} 
\includegraphics[width=3.2cm]{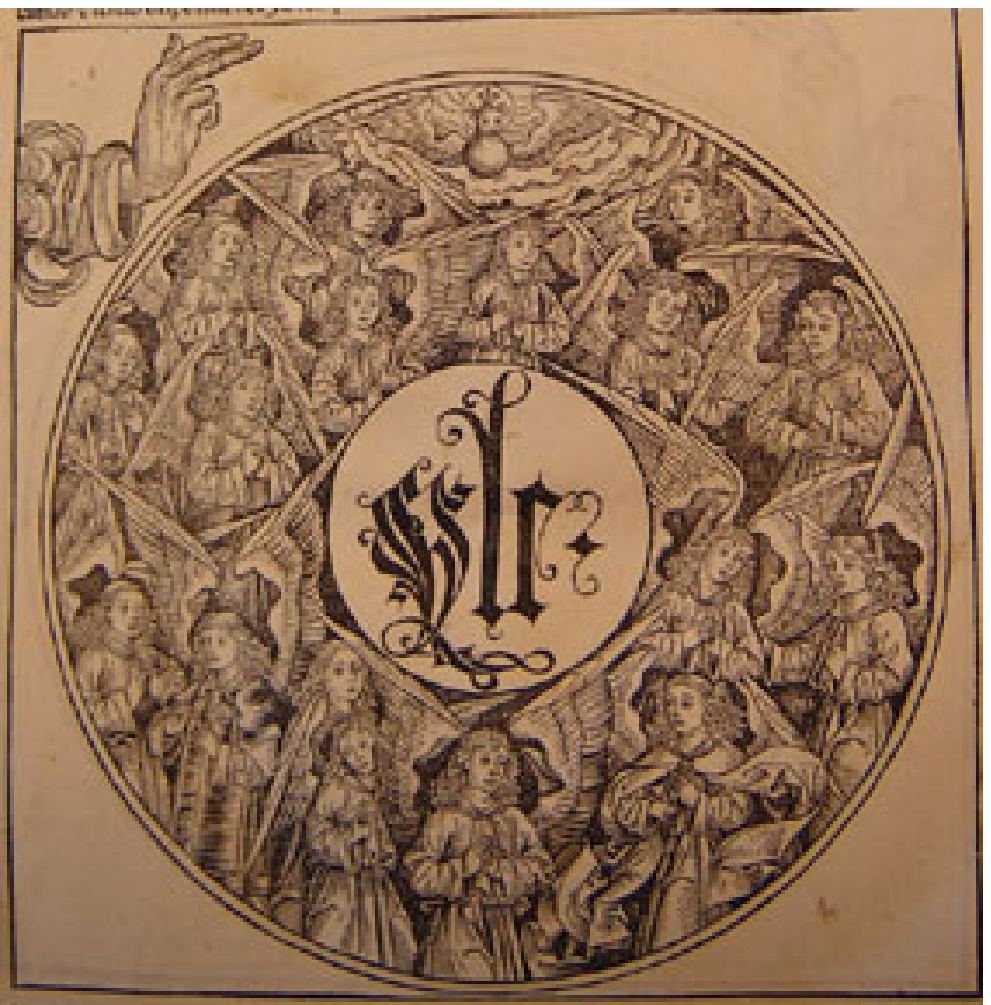}
\includegraphics[width=3.2cm]{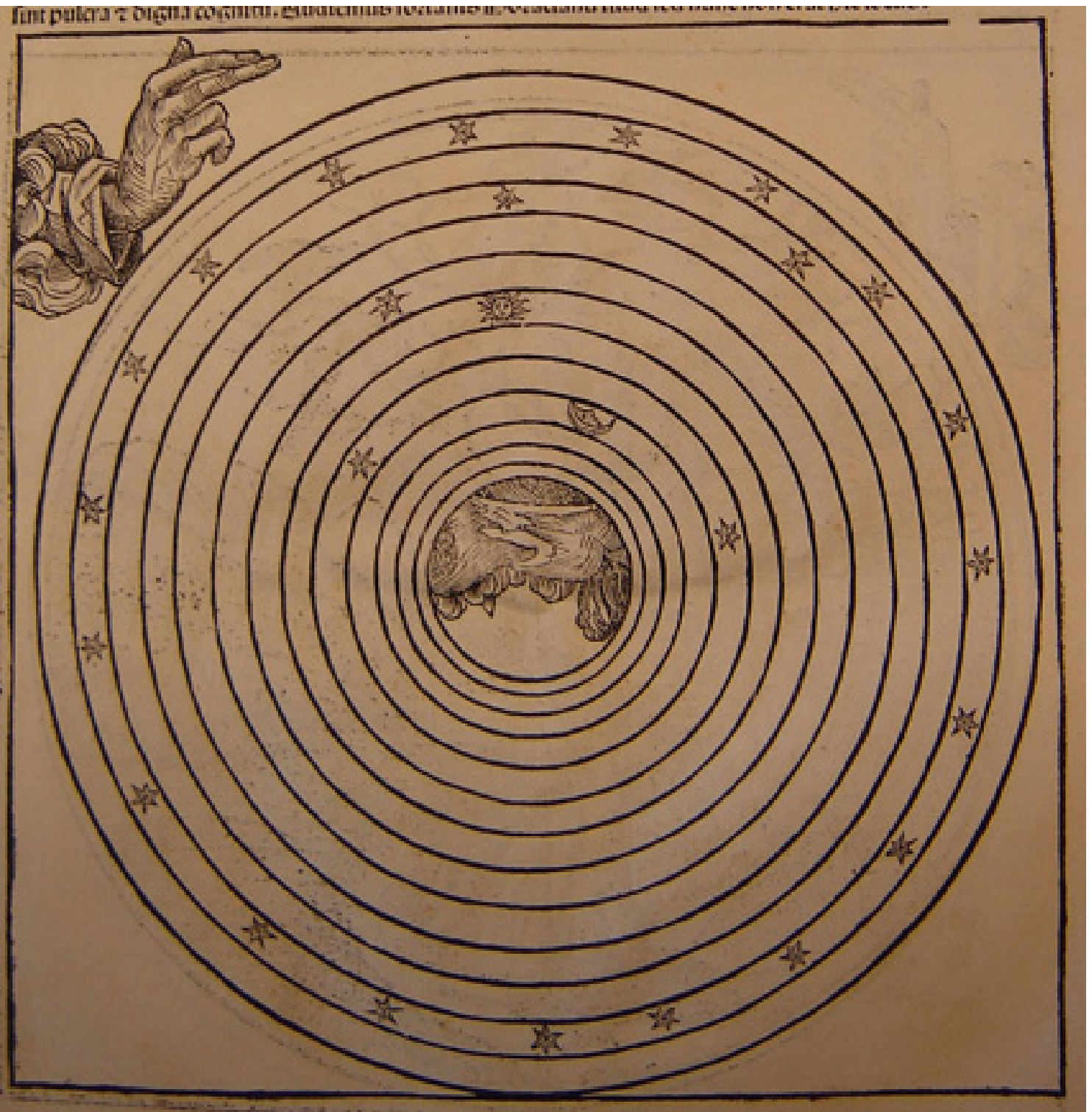}
\includegraphics[width=3.2cm]{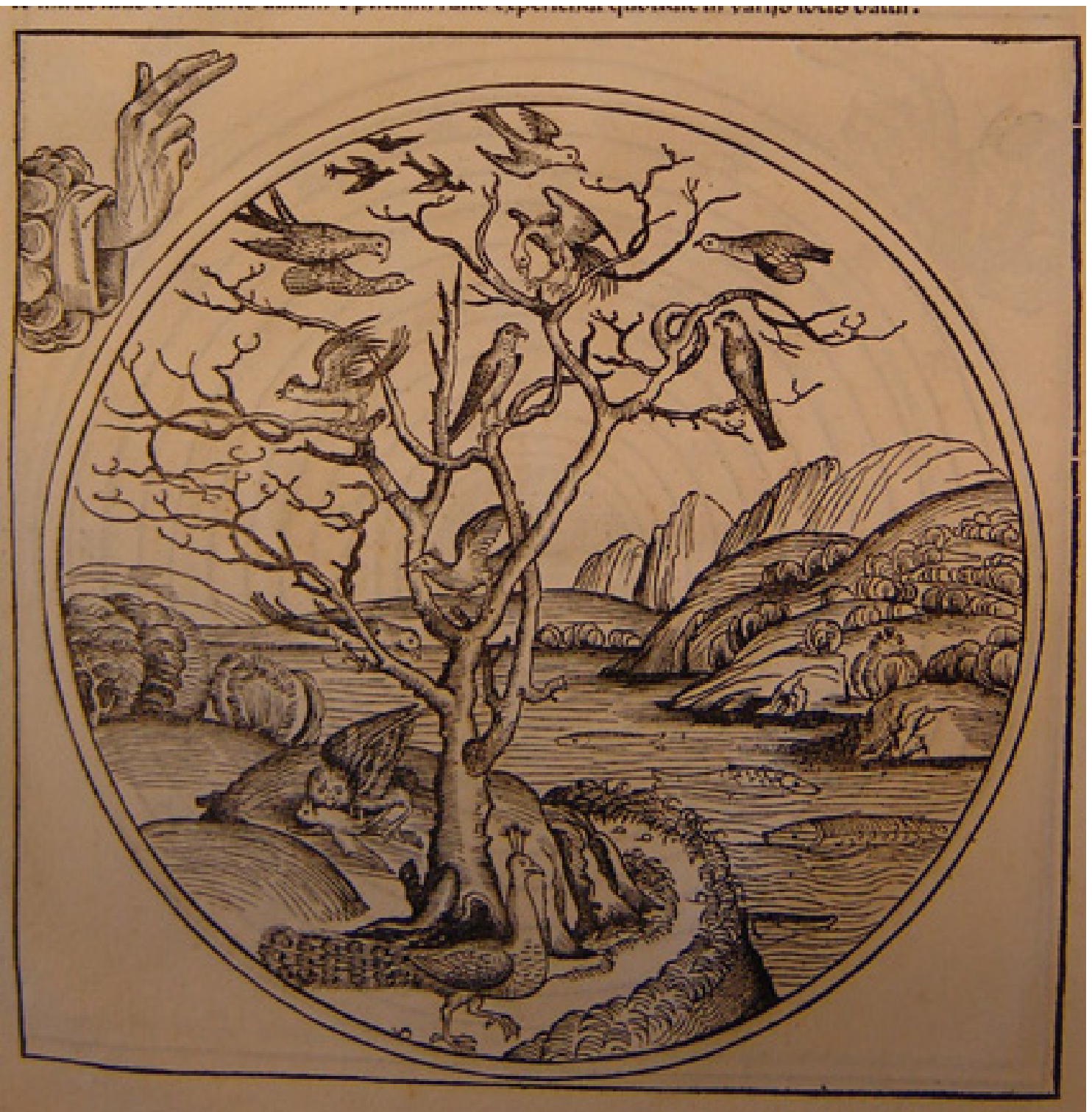}
\includegraphics[width=3.3cm]{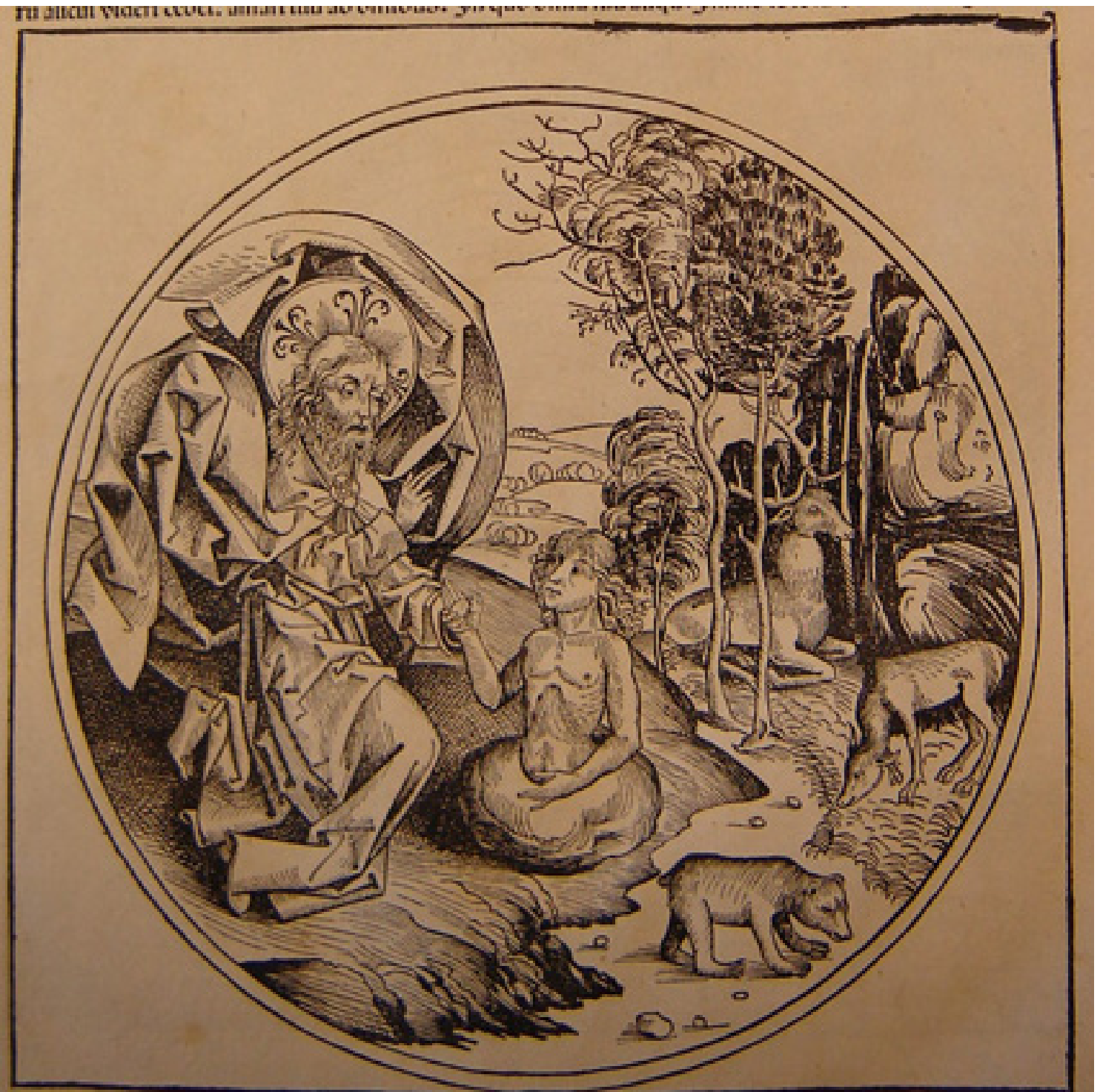}
\caption{{\it In principio...} the N\"urnberg chronicle, by Hartmann Schedel, 1493.} \label{fig1}\end{center}\end{figure}

This original Latin edition includes, among its nearly 1800 engravings, and of most interest to us, a thorough description of the seven
ages of the world after Creation, beginning with a biblical heptameron, of which already the forth day shows a nice geocentric Ptolemy's
universe (second picture in Fig.\,\ref{fig1}).

The Library also owns a Venice 1484 copy of Dante's Divine Comedy, with comments by Cristoforo Landino, which beautifully illustrates
Beatrice guiding the pilgrim across the astronomical and metaphysical spheres of the celestial Paradise. 

\begin{figure}[h!tb]
\begin{center}
\includegraphics[width=11cm]{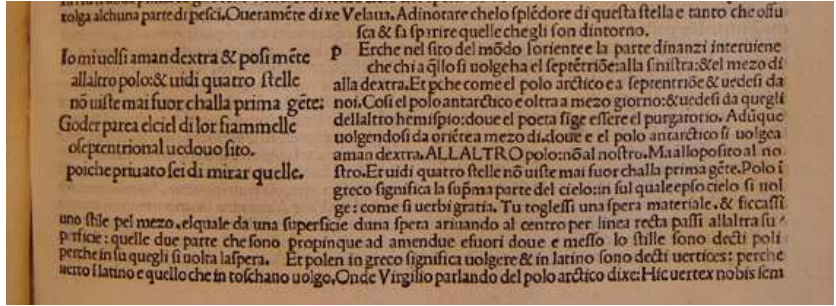}
\caption{Dante's Divine Comedy (Venice, 1484) with comments by Cristoforo Landino. 
{\it 
To the right hand I turned, and fixed my mind/ Upon the other pole, and saw four stars/ Ne'er seen before save by the primal people./ 
Rejoicing in their flamelets seemed the heaven./ O thou septentrional and widowed site,/ Because thou art deprived of seeing these!}
(Purgatorio I, 22-27, translation by H.W.Longfellow). ``{\it Quatro stelle...}'': is he making reference to the Southern Cross?
(\cite[Gangui 2008]{Gangui2008}).} 
\label{fig2}\end{center}\end{figure} 


These and dozens of other equally interesting books build up an important part of both literary and pre-scientific culture. However, only
a handful of persons (mainly researchers) have access to them. Just as we all know that the feeling of seeing Mars in pictures cannot be
compared to the experience of actually seeing it through a telescope, in the case of old books, the project of exhibiting them is much
more rewarding than just looking at them through the internet. Partly because of this, we are organizing an exhibition of rare books with
astronomical flavor in years 2009 -- 2010. 

\begin{figure}[h!tb]
\begin{center}
\includegraphics[height=4cm]{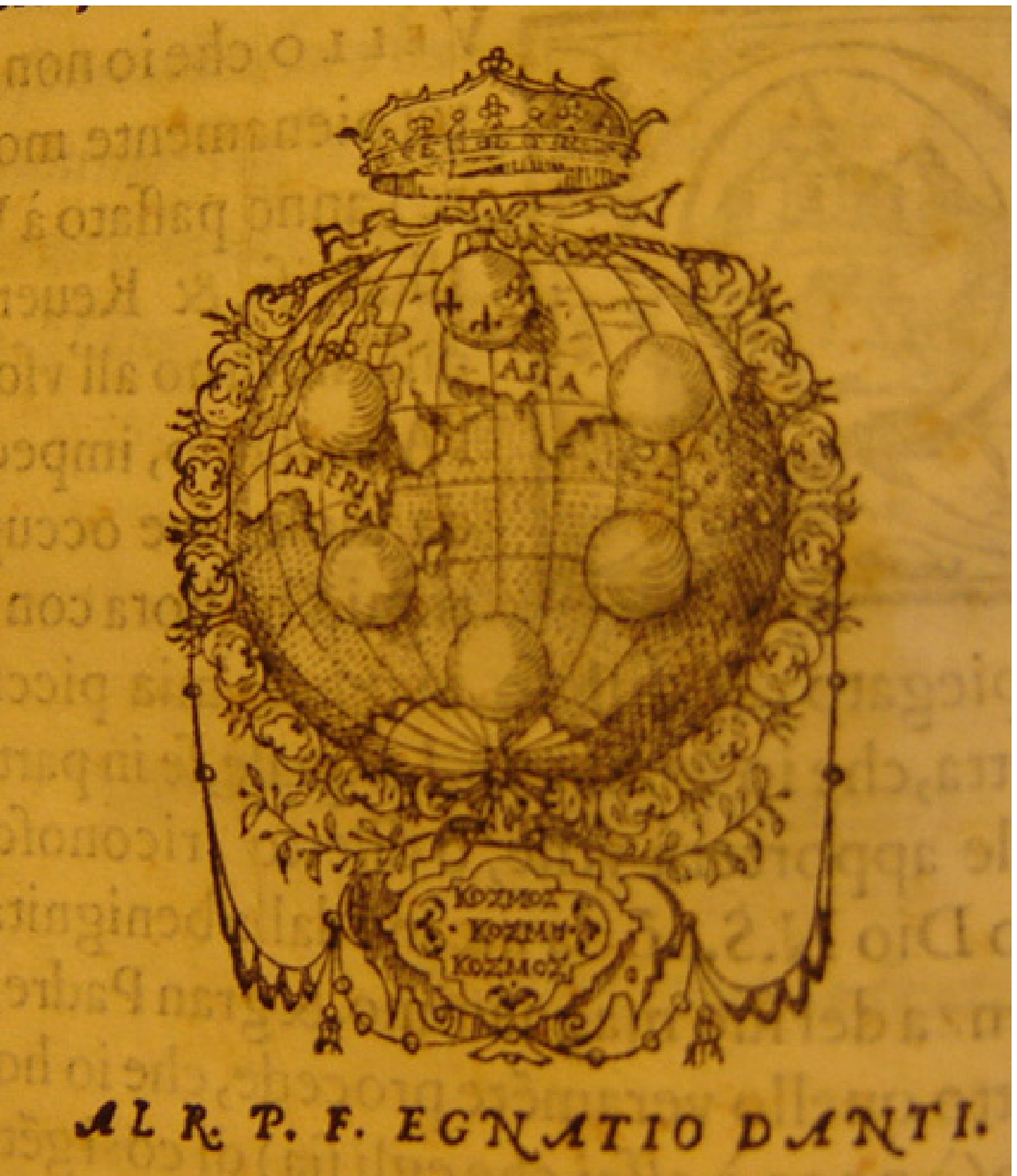}
\includegraphics[height=4cm]{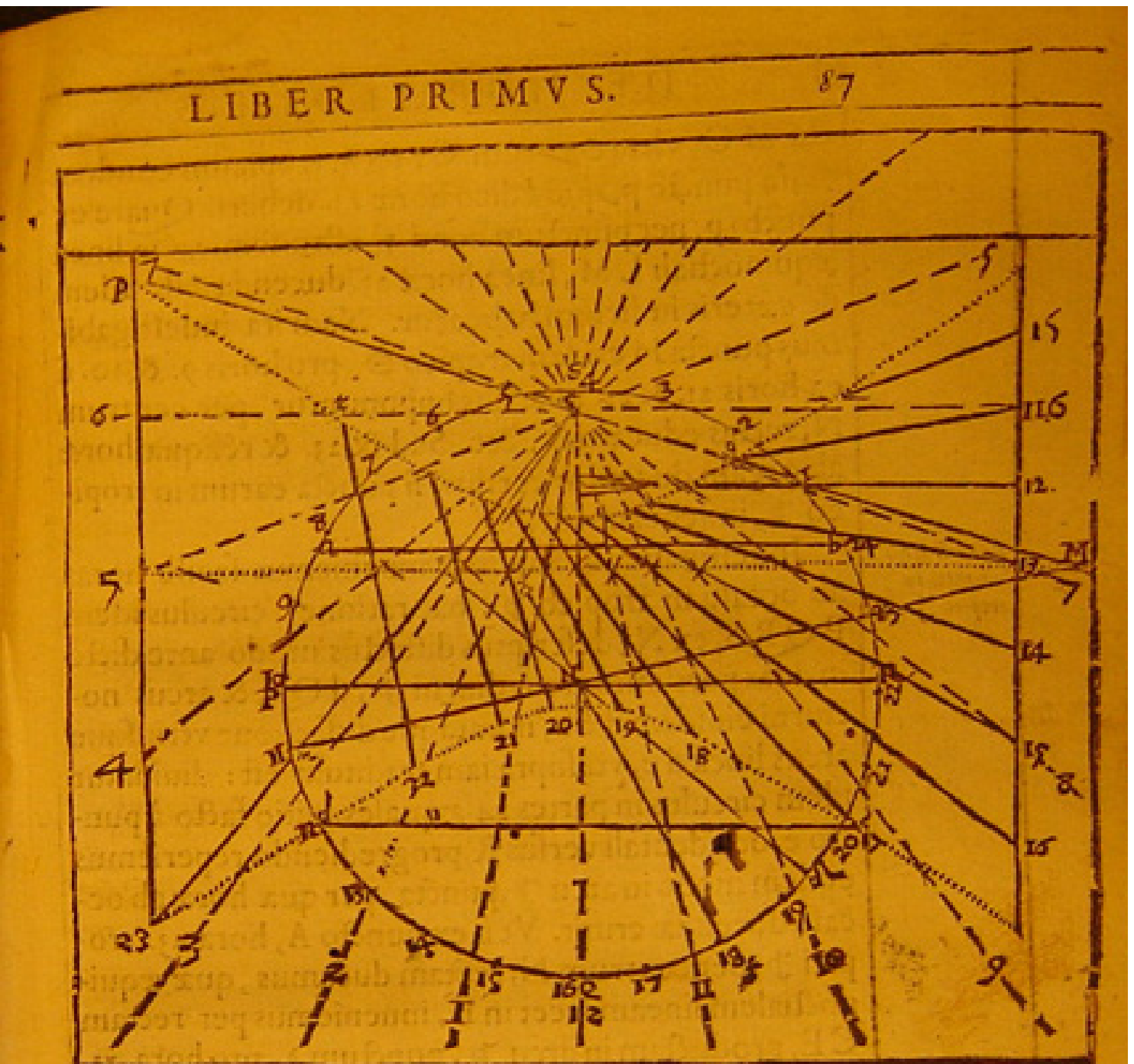} \\
\includegraphics[height=3.8cm]{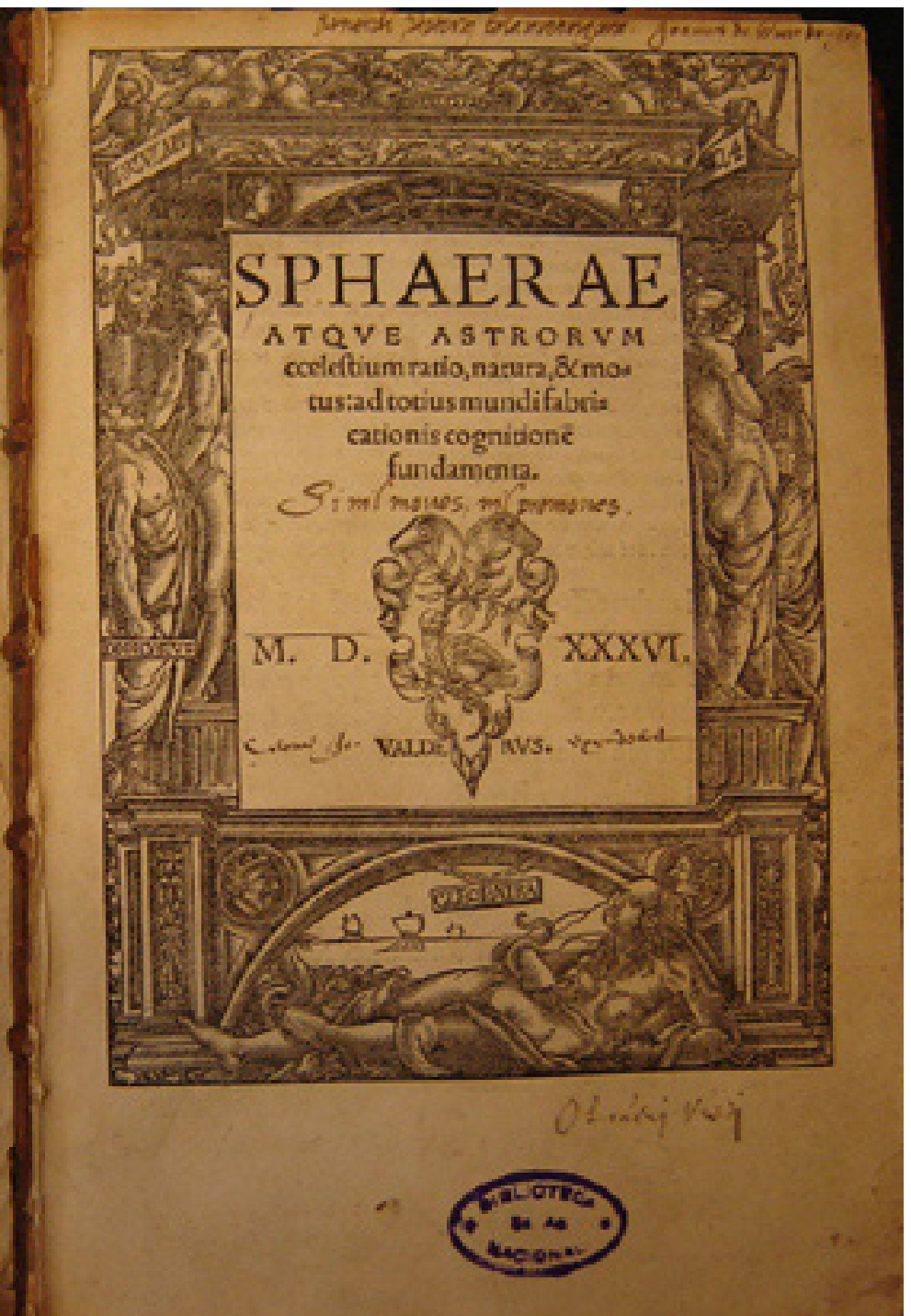}
\includegraphics[height=3.8cm]{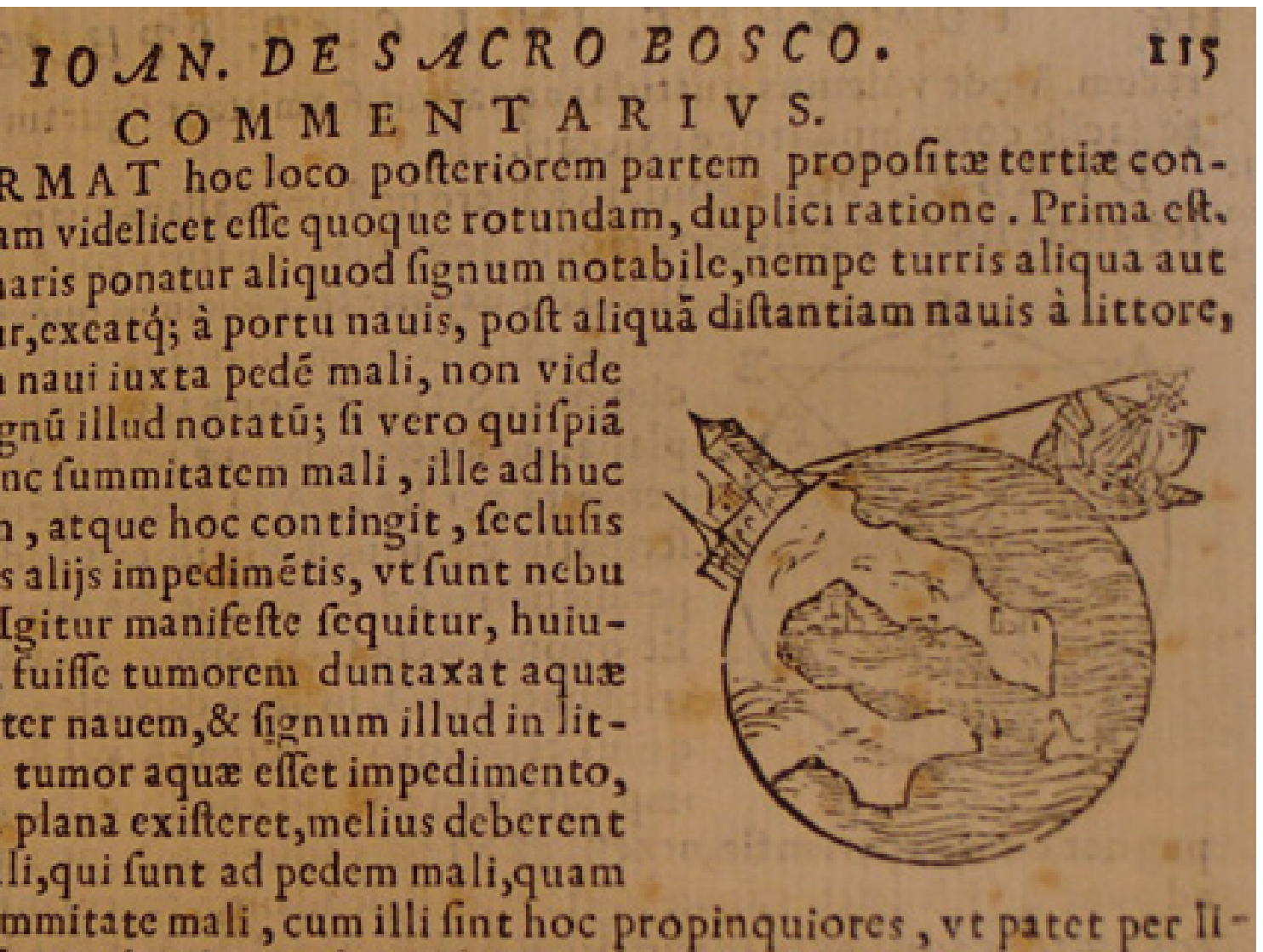}
\caption{Egnatio Danti's Trattato dell'astrolabio (1569), 
Voello's De horologiis (1608), 
Iacobus Valderus' Sphaera (1536), 
Clavius' commentary of Sacro Bosco's Sphaera (1585).} 
\label{fig3}\end{center}\end{figure}

Many other old or rare books are also included in our list: Iacobus Valderus' Sphaera (1536), Egnatio Danti's Trattato dell'astrolabio
(1569), Clavius' commentary of Sacro Bosco's Sphaera (1585), Joanne Voello's De horologiis (1608), Henrico Hofmanno's De octantis (1612),
Blaev's Theatrum orbis terrarium (1640), and a long etcetera. Some of these books were presumably brought to the River Plate by the
Catholic religious order of the Jesuits during the XVIII century. Hence, these books allow us to reconstruct the kind of science imparted
in our country at that time.

\begin{figure}[h!tb]
\begin{center}
\includegraphics[height=5.1cm]{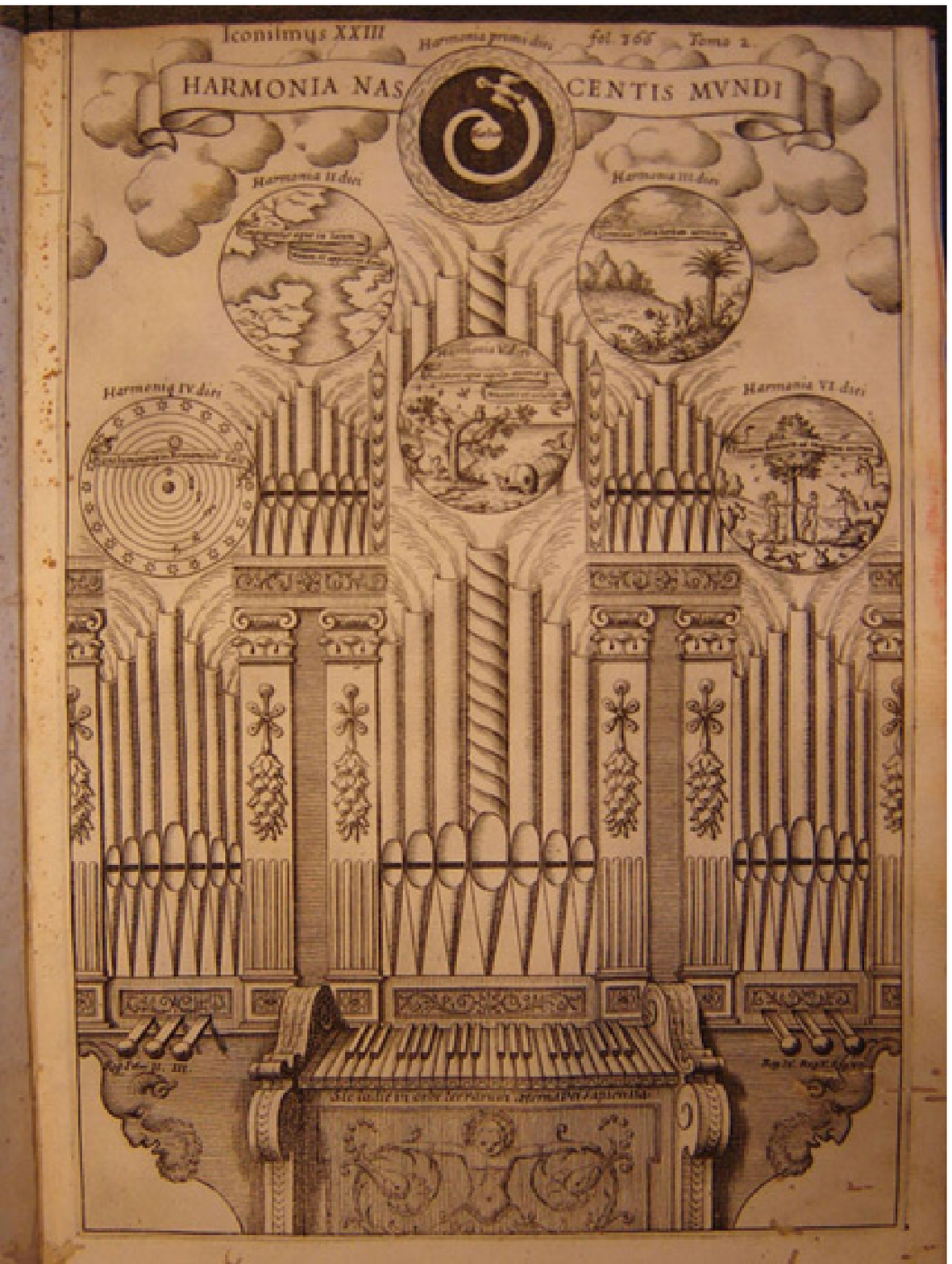}
\includegraphics[height=5.1cm]{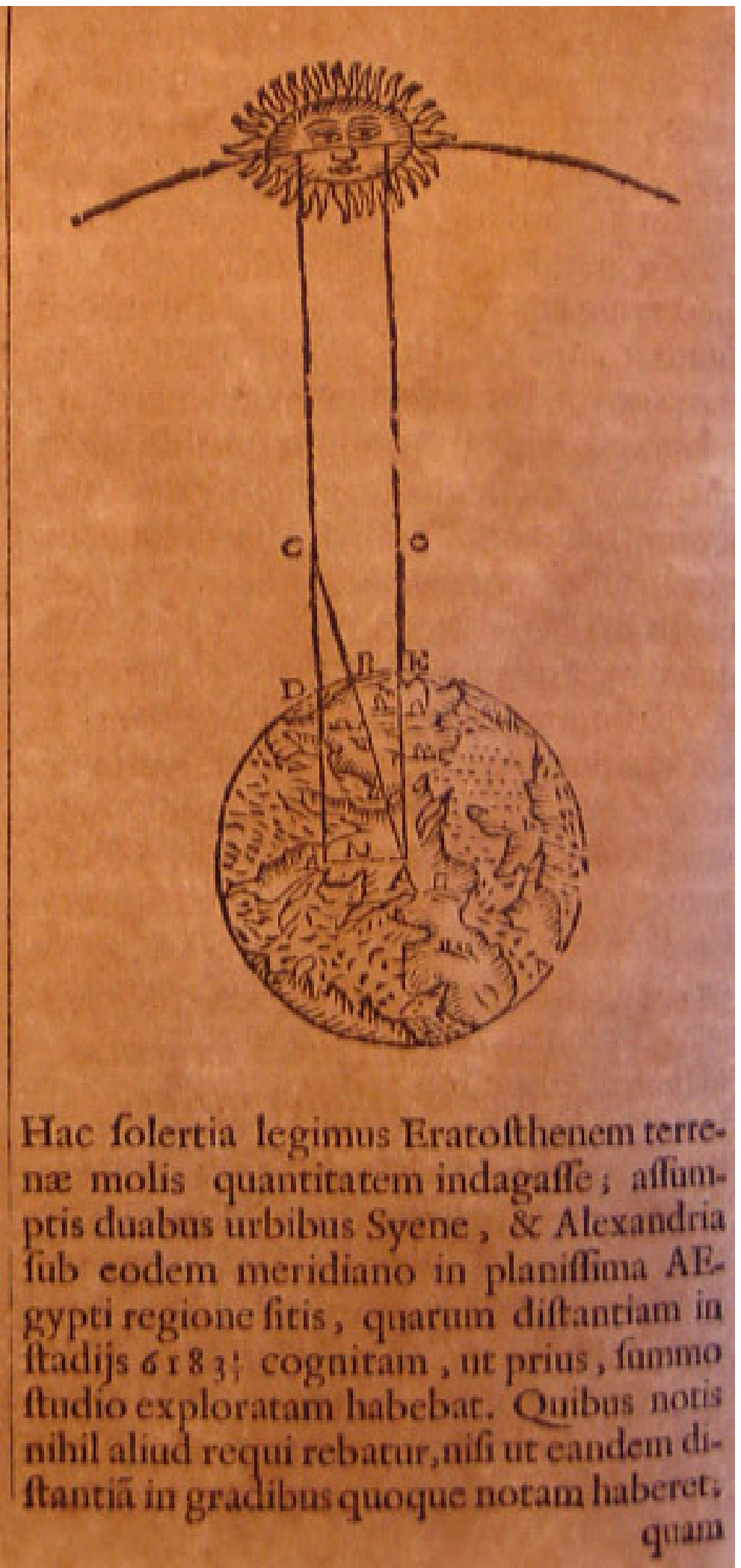}
\includegraphics[height=5.1cm]{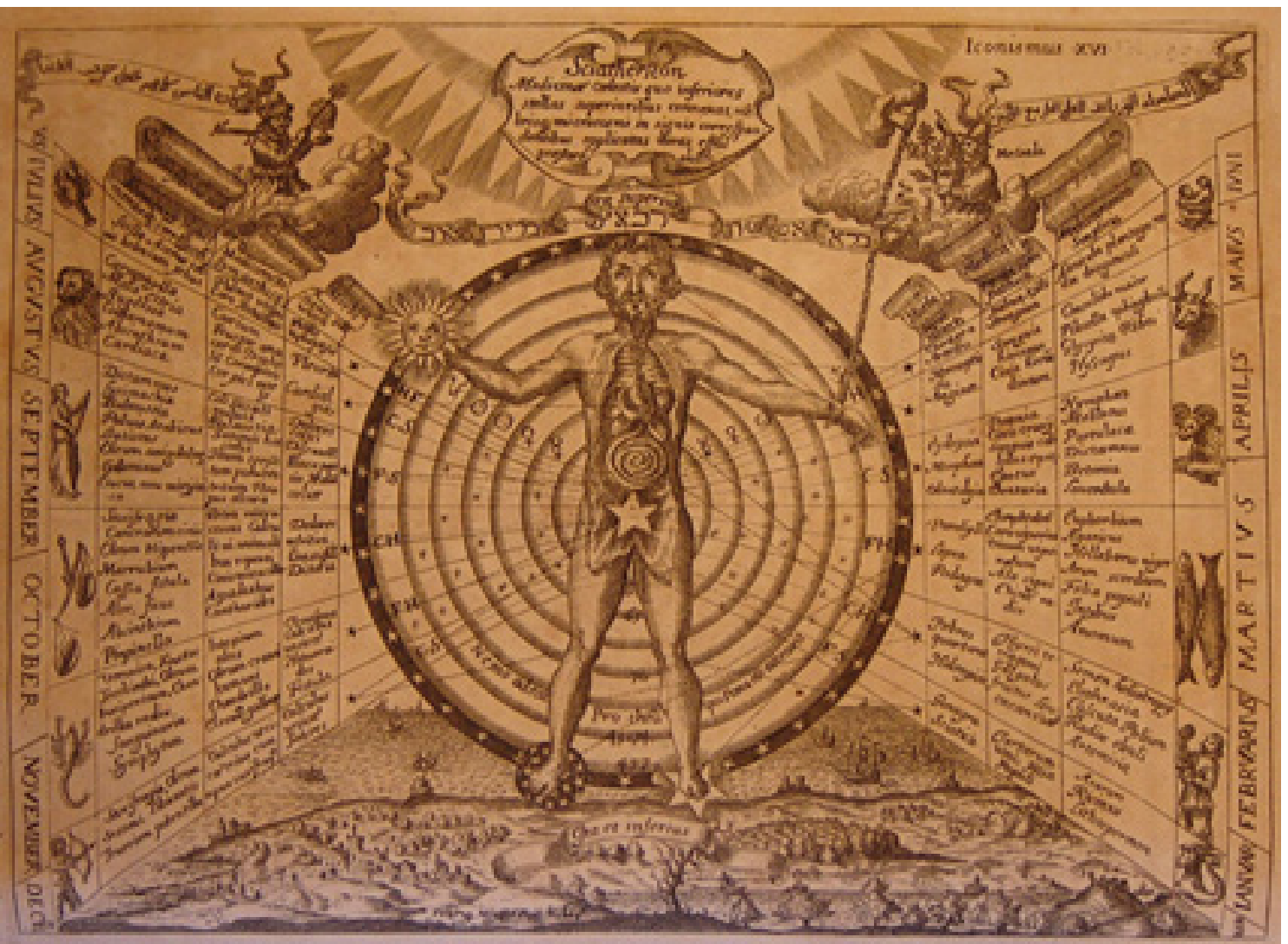}
\caption{Athanasius Kircher's Musurgia universalis, 1654 (left) and a couple of images from Ars Magna Lucis et Umbrae, 1671 
(depicting Eratosthenes' method to measure the circumference of the earth -image in the middle-  
 and the micro and macro cosmos connection in a sort of Vitruvian man representation, on the right).}
\label{fig4}\end{center}\end{figure}

Among these books, German Jesuit Athanasius Kircher's thick and numerous volumes, with their gorgeous engravings illustrating all
possible areas of knowledge, naturally attract the attention: his Musurgia universalis (1654) depicts cosmic harmony as musical sounds
emanating from an organ played by God; his Ars Magna Lucis et Umbrae (1671) shows the analogy between micro and macro cosmos, with man
placed both at the center of the universe and of the zodiacal signs (Fig.\,\ref{fig4}). 

In Kircher's Mundus subterraneus (1678) the frontispiece shows a lady (the allegory of Astronomy) inspecting a celestial globe and taking
notes with a {\it plume d'oiseau}, while another feminine character looks through a telescope. Inside the book we find a sketch of the
Moon as an aqueous body, with spots, mountains and sources, as well as other rough earthy textures scattered on its visible face, and
another of the Sun, divided in different regions, including an equatorial torrid zone, quite similar to the one of the Earth at the time,
and covered with drawings of smoke and fires as sources of Sun spots.

\begin{figure}[h!tb]
\begin{center}
\includegraphics[height=3.7cm]{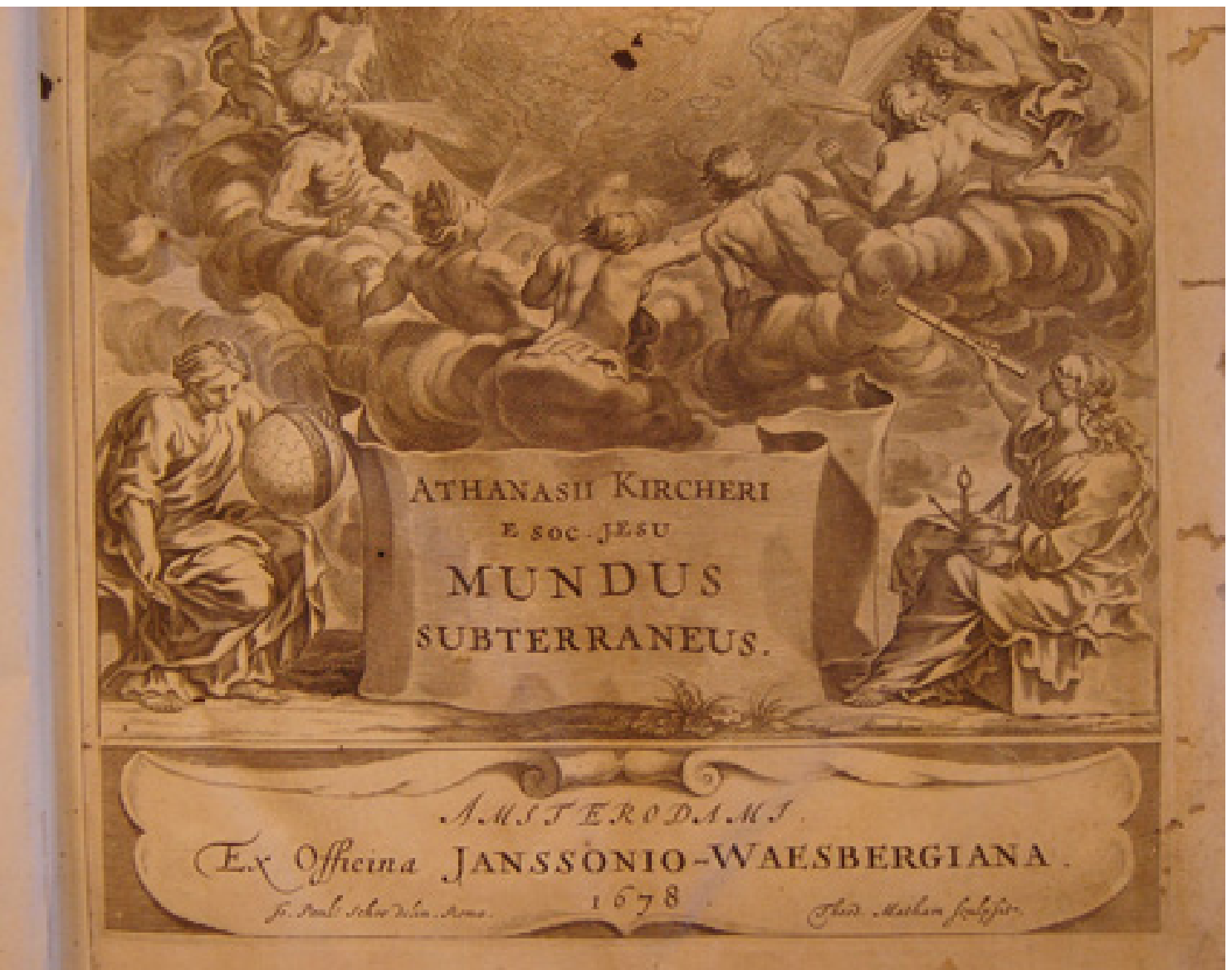}
\includegraphics[height=3.7cm]{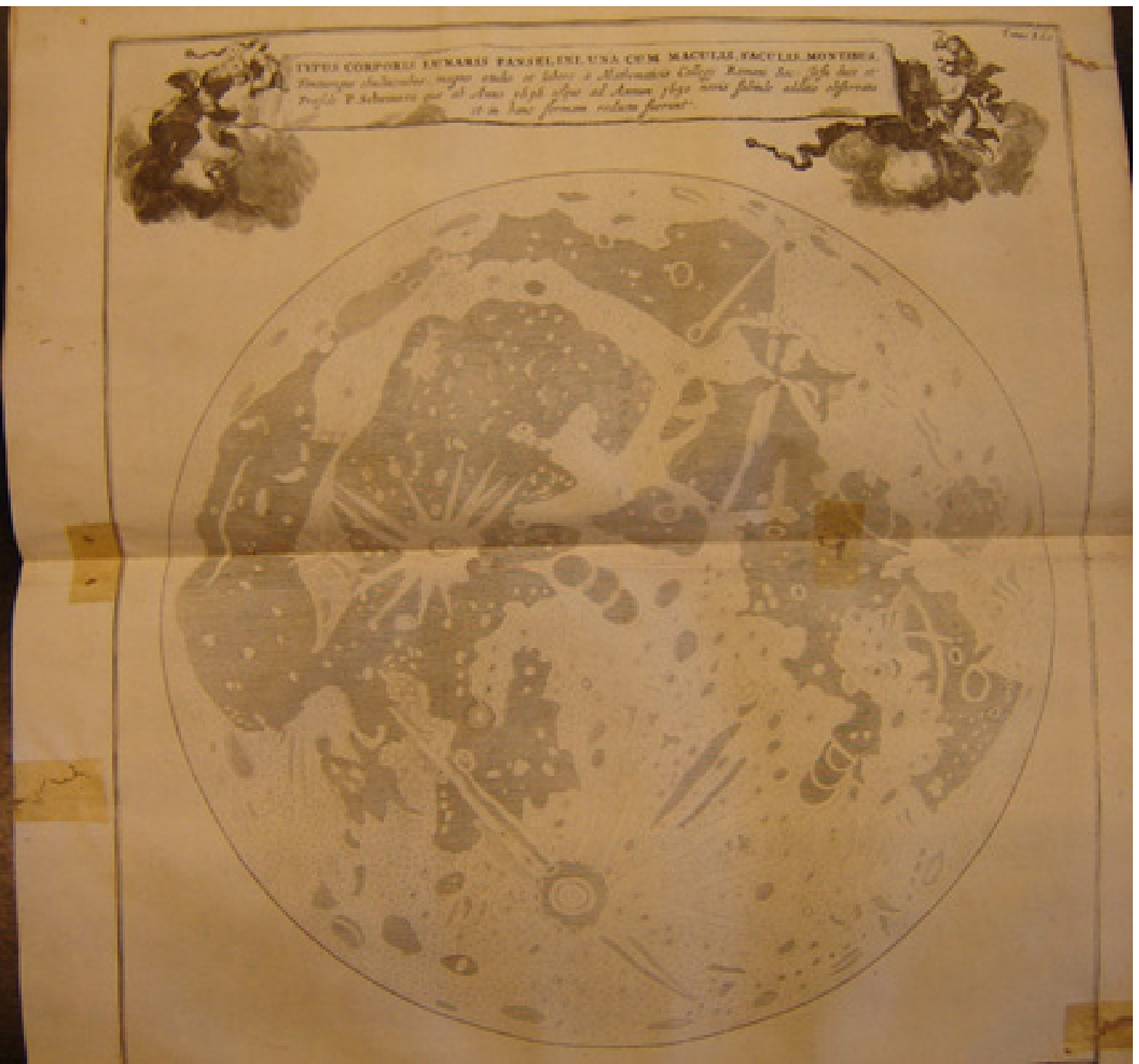}
\includegraphics[height=3.7cm]{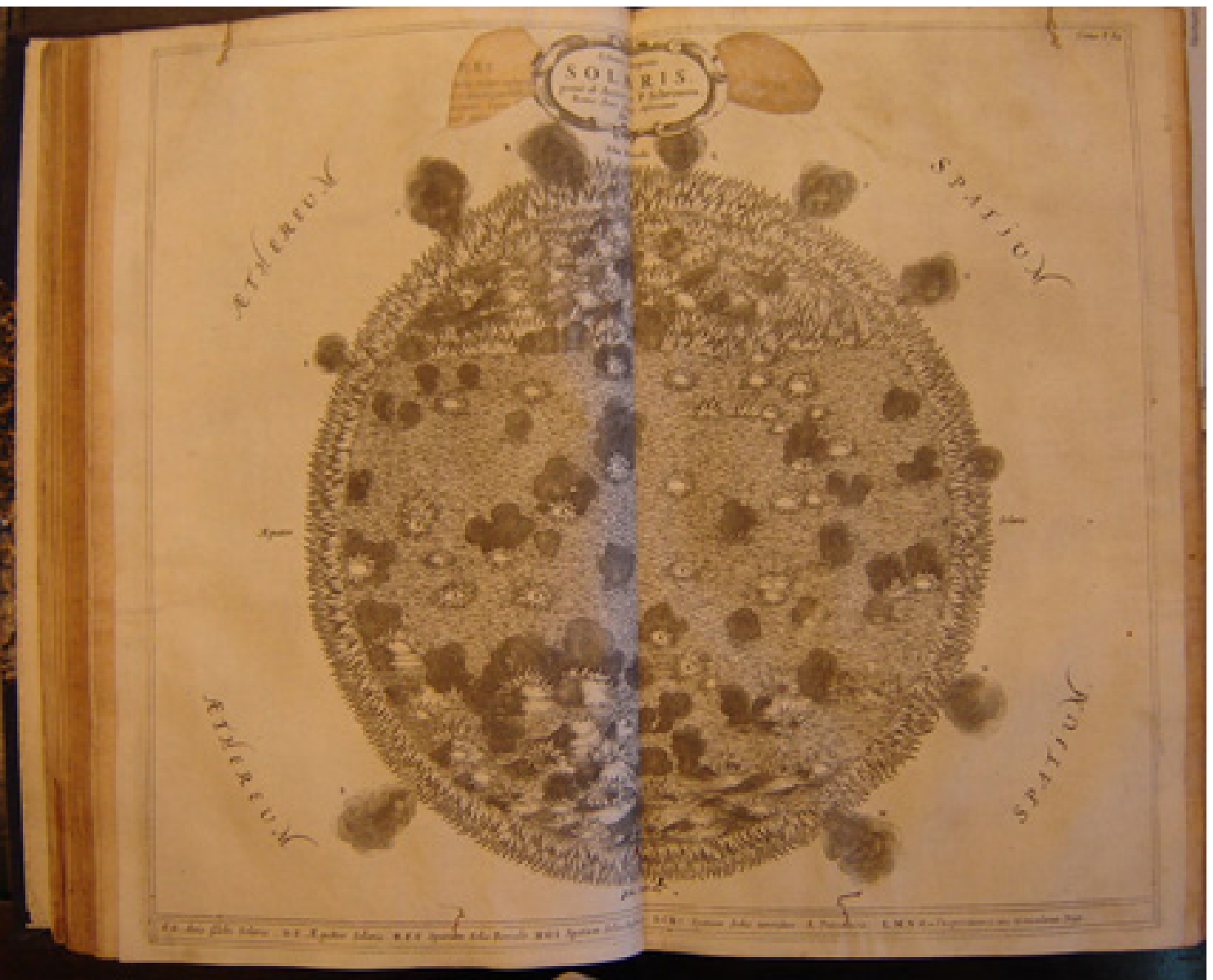}
\caption{Kircher's Mundus subterraneus (1678). Frontispiece and two sketches: of the Moon (middle) and of the Sun (right).}
\label{fig5}\end{center}\end{figure}

\begin{figure}[h!tb]
\begin{center}
\includegraphics[height=6cm]{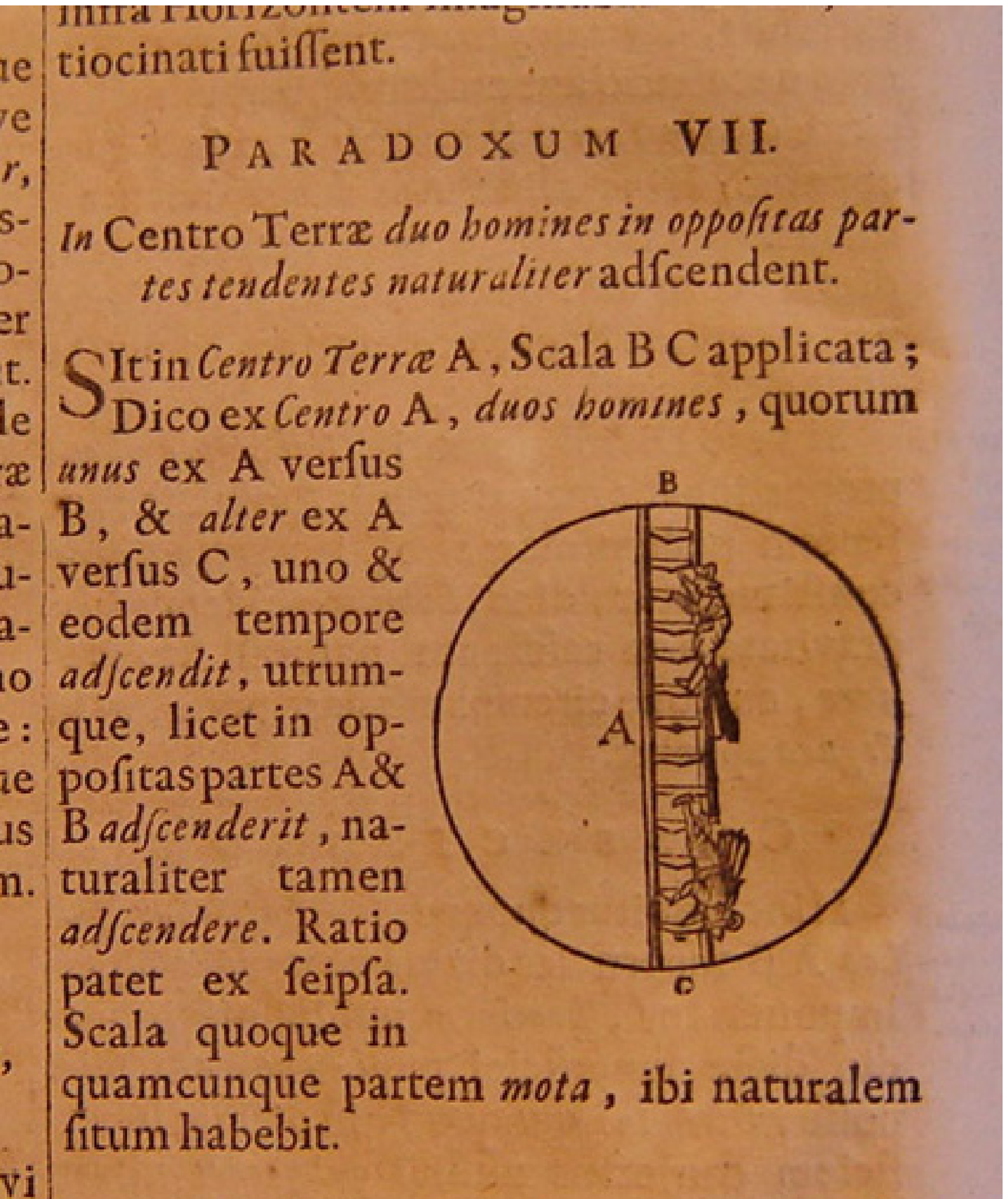}
\includegraphics[height=6cm]{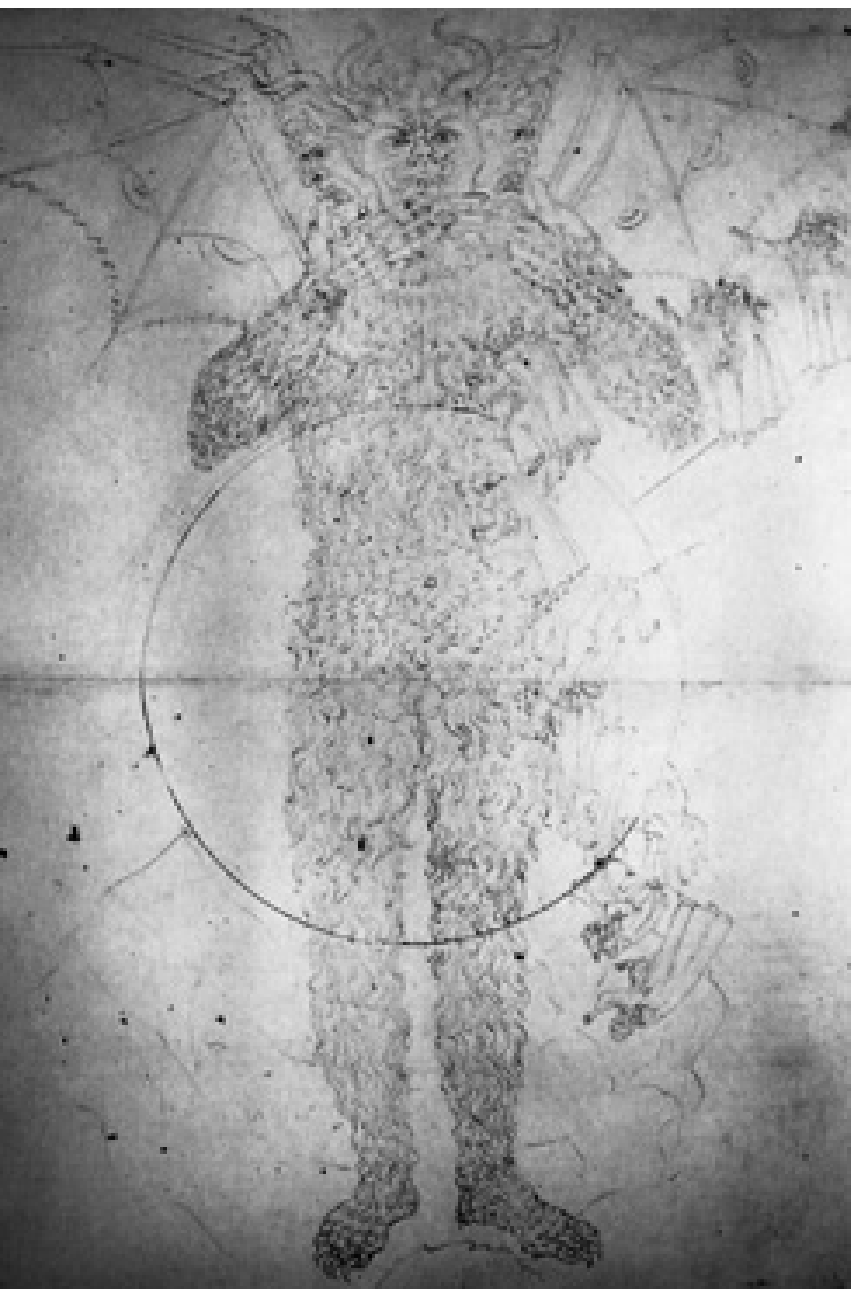}
\caption{The image on the left is from Kircher's Mundus subterraneus and shows how two persons travelling from point A --located at the
  center of the Earth-- have to go in opposite directions to reach the surface, ``climbing upwards'' with their stairs. A similar
  situation occurred to Dante in the last canto of the Inferno when, together with Virgilio, descended along the giant Lucifer's body,
  who was stuck in the center of the Earth and, suddenly, got turned upside down, as shown in this Sandro Botticelli's drawing for the
  {\it Commedia} (image on the right).}
\label{fig5}\end{center}\end{figure}

Finally, father Gaspare Schotto's Iter exstaticum Kircherianum, of 1671, shows in its very frontispiece a peculiar engraving of Kircher
himself (as Theodidactus, the disciple of God) when, guided by angel Cosmiel, he travels across the universe, in a clear parallel to
Dante's voyage following his {\it donna-angelo}. The universe depicted is neither Ptolemaic nor Copernican, but that of Tycho Brahe, with
the Sun orbiting the Earth, while the rest of the planets complete their movements around the Sun; a clear eclectic cosmology agreeing
well with the author's world view.

\begin{figure}[h!tb]
\begin{center}
\includegraphics[height=5.5cm]{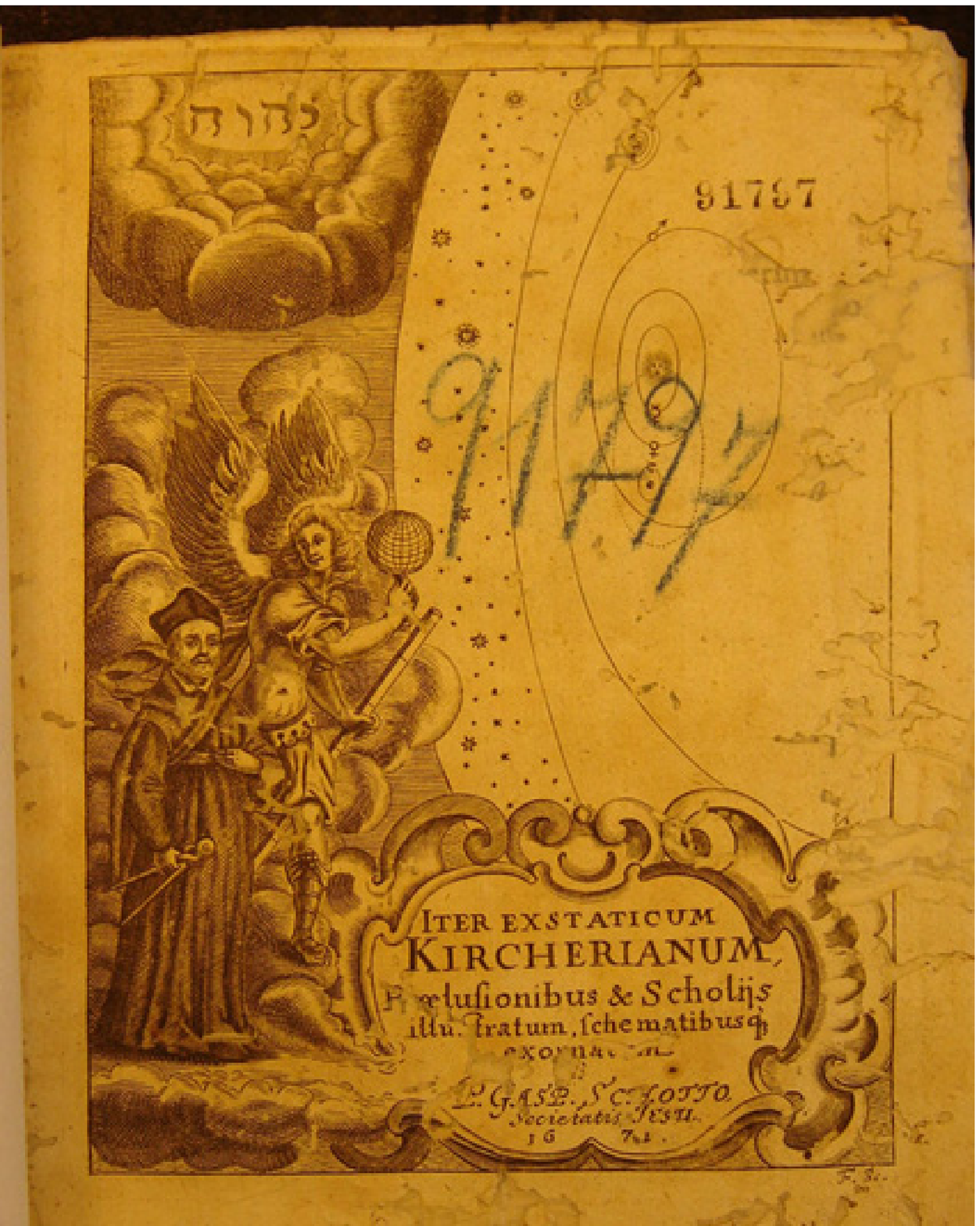}
\includegraphics[height=5.5cm]{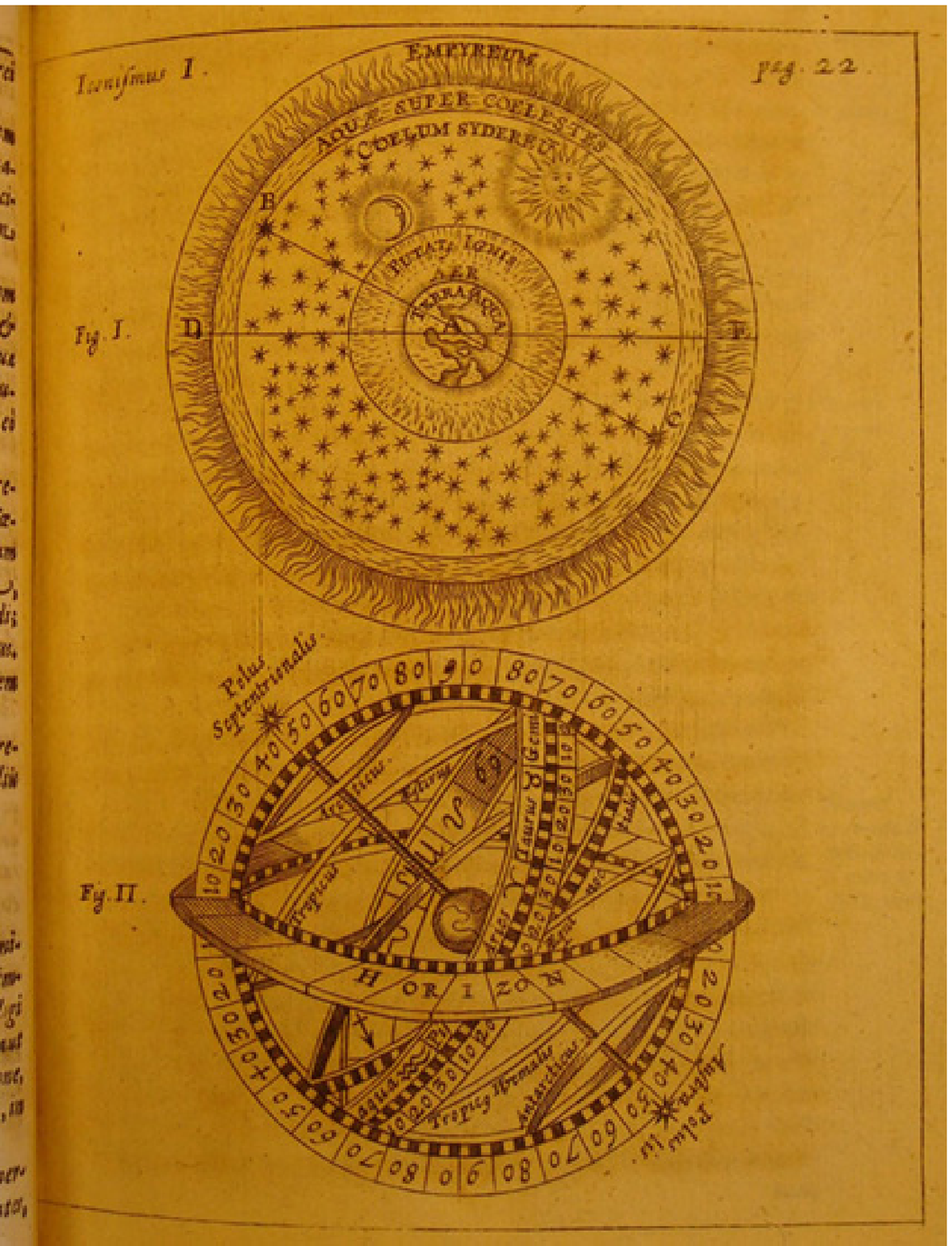}
\includegraphics[height=5.5cm]{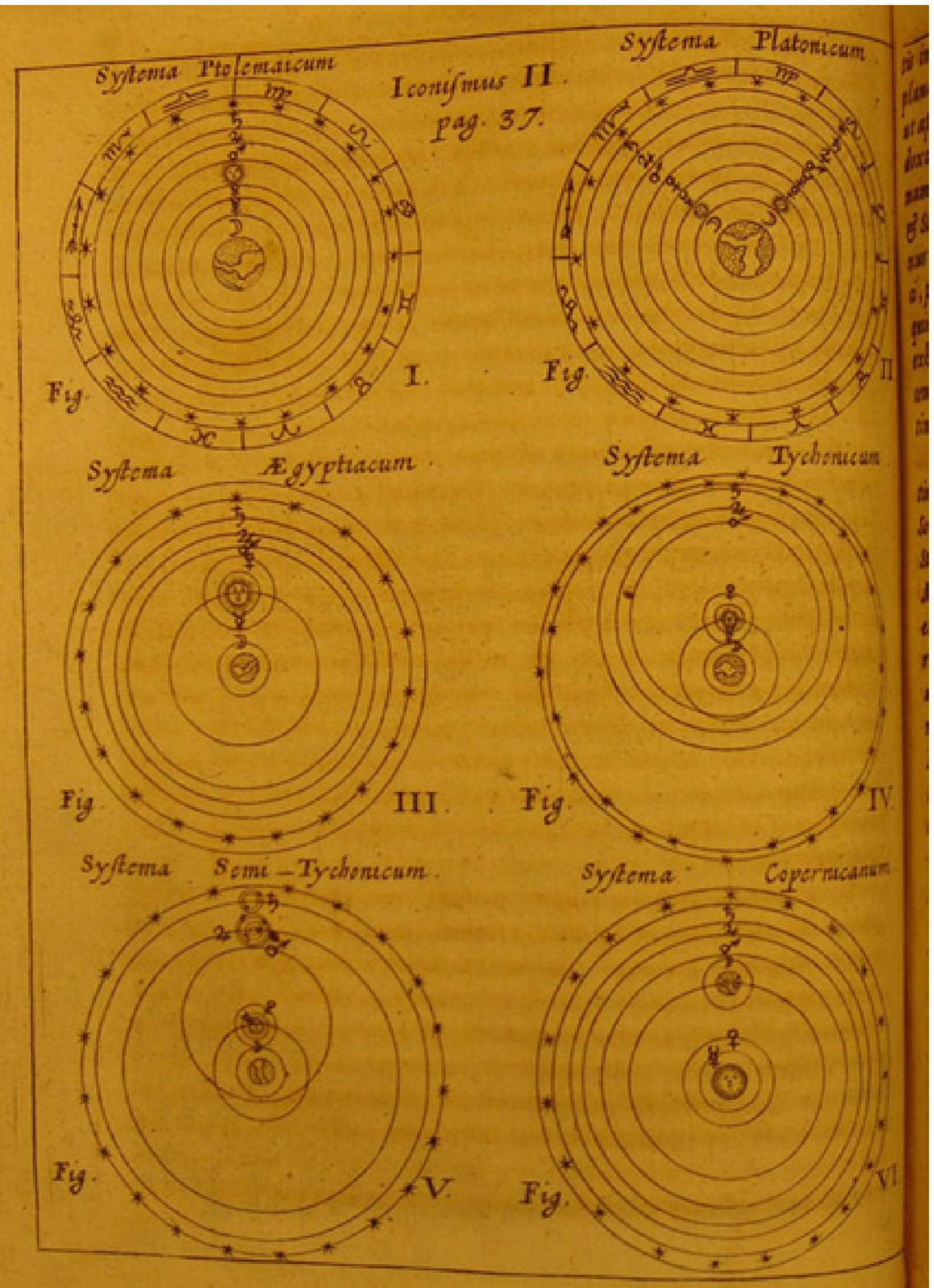}
\caption{Gaspare Schotto's Iter exstaticum Kircherianum, 1671. Frontispiece depicting a Tychonic universe (left) and two of the most
  noteworthy pages: a representation of the elements surrounding the Earth, from the Empyrean downwards, and the different principal
  circles of the sky (middle), while the image on the right shows a collection of the most representative systems of the world, from the
  Ptolemaic to the Copernican models.}
\label{fig6}\end{center}\end{figure}



The planned exhibition will collect not only these and other books, but also historical documents, maps and drawings (may be also
artifacts). Hopefully, it will offer a timeline of our understanding of old and renaissance astronomy and, with it, part of the 
{\it imago mundi} of the time. 

\vspace*{-0.5 cm}

\end{document}